\documentclass[aps,nofootinbib,notitlepage,pra]{revtex4-1}
\pdfoutput=1

\usepackage[utf8]{inputenc}
\usepackage[T1]{fontenc}
\usepackage{mathpazo}
\usepackage{amsmath}
\usepackage{mathtools}
\usepackage{amsfonts}
\usepackage{amssymb}
\usepackage{amstext}
\usepackage{hyperref}
\usepackage{graphicx}
\usepackage{xcolor}

\DeclareMathOperator{\tr}{tr}
\DeclareMathOperator{\rk}{rank}
\newcommand{\ee}{\ensuremath{\mathrm{e}}}
\newcommand{\nn}{\nonumber}

\definecolor{darkred}{rgb}{0.9,0,0}
\definecolor{darkgreen}{rgb}{0,0.9,0}
\definecolor{darkblue}{rgb}{0,0,0.9}

\hypersetup{
colorlinks,
linkcolor=darkblue,
filecolor=darkgreen,
urlcolor=darkblue,
citecolor=blue}

\begin{document}
\title{Informational work storage in quantum thermodynamics}
\author{Shang-Yung Wang}
%\email{sywang@mail.tku.edu.tw}
\affiliation{Department of Physics, Tamkang University, New Taipei City 25137, Taiwan}

\begin{abstract}
We present a critical examination of the difficulties with the quantum versions of a lifted weight that are widely used as work storage systems in quantum thermodynamics. To overcome those difficulties, we turn to the strong connections between information and thermodynamics illuminated by Szilard's engine and Landauer's principle, and consider the concept of informational work storage. This concept is in sharp contrast with the usual one of mechanical work storage underlying the idealization of a quantum weight. An informational work storage system based on maximally mixed qubits that does not act as an entropy sink and is capable of truly distinguishing work from heat is studied. Applying it to the problem of single-shot work extraction in various extraction schemes, we show that for a given system state the maximum extractable work is independent of extraction scheme, in accordance with the second law of thermodynamics.
\end{abstract}

%\keyword{quantum thermodynamics; work extraction and storage; thermodynamic resource theory}

\maketitle

\section{Introduction}

The strong connections between information and thermodynamics, first conceived by Maxwell~\cite{MaxwellDemon,Rex:2002,Maruyama:2009} and later quantified by Szilard~\cite{Szilard:1929}, Shannon~\cite{Shannon:1948}, and Landauer~\cite{Landauer:1961}, have continued to provide deep insights into the interplay between information theory and thermodynamics of quantum systems~\cite{Parrondo:2015,Goold:2016}.
Inspired by the resource theory of entanglement in quantum information~\cite{Horodecki:2009}, quantum thermodynamics has recently been formulated as a resource theory~\cite{Janzing:2000,Gour:2015,Brandao:2013}. 
A powerful framework for thermodynamic resource theory is that of thermal states at temperature $T$ and energy-conserving transformations, also known as thermal operations~\cite{Brandao:2013}.
The hallmark of thermal operations lies in their abstraction and generality, in that any thermal states at temperature $T$ associated with any Hamiltonians of the system, heat bath, and auxiliary systems are free; any energy-conserving global unitary transformations are allowed operations; and any states of the system, heat bath, and auxiliary systems that are not thermal states at temperature $T$ are resources. 
Many interesting results and their information-theoretic connections have been obtained using thermal operations (see, e.g., Refs.~\cite{Lostaglio:2015,Brandao:2015,Horodecki:2013}). 

A fundamental difference between the theory of entanglement and quantum thermodynamics is that the latter concerns not only state transformations but also energy transfer and heat to work conversion.
To consistently account for energy conservation, heat to work conversion, and the extraction or expenditure of work in thermodynamic processes, a work storage system has to be explicitly included~\cite{Horodecki:2013,Kosloff:2014}.
The most widely used work storage systems in thermal operations are quantum versions of a lifted weight that can be raised or lowered when work is done on or by it~\cite{Horodecki:2013,Skrzypczyk:2014}. 
The underlying notion is that work can be thought of as the ability to increase the \emph{energy content} of a work storage system, and hence can be analogously defined as the change in the energy of a quantum weight.
However, quantum weights do not seem to fit naturally into thermal operations. Specifically, quantum weights require specifically tailored Hamiltonians, they are prepared in certain pure states instead of thermal states at temperature $T$, and their inclusion imposes an additional restriction on the class of allowed energy-conserving global unitaries~\cite{Horodecki:2013,Skrzypczyk:2014,Gemmer:2015,Richens:2016}.  

Recently, work extraction in quantum thermodynamics~\cite{Horodecki:2013,Skrzypczyk:2014} has been generalized to cases where the quantum weight transfers from a single energy level to a range of multiple energy levels~\cite{Gemmer:2015}, and cases where the extracted work is allowed to fluctuate freely~\cite{Richens:2016}. 
In particular, the maximum extractable work in the general case considered in Ref.~\cite{Gemmer:2015} contains an additional contribution that is independent of the initial system state, and increases without bound as the density of states of the quantum weight increases. This result leads to a conceptual inconsistency that signals a violation of the second law of thermodynamics. 
The inconsistency was later resolved by the present author~\cite{Wang:2017} by recognizing that the additional contribution is heat-like rather than work-like, and should count as heat transferred from the heat bath to the quantum weight instead of as work extracted from the system. 
Hence the maximum extractable work from a system state remains unchanged for the general case considered in Ref.~\cite{Gemmer:2015}, and there is no violation of the second law. 
Moreover, it has been shown~\cite{Wang:2017} that the origin of the inconsistency is that when the extracted work is allowed to fluctuate, the quantum weight does not truly distinguish between the work extracted to it from the system and the heat transferred to it from the heat bath. 
This in turn calls into question the naive notion that work in quantum thermodynamics can be reliably quantified as the change in the energy of a quantum weight. 

The suitability of using quantum weights as work storage systems in thermal operations, and their capability of truly distinguishing work from heat are interesting open questions in quantum thermodynamics.
In this article, we present a critical examination of the quantum weight, and provide a detailed account of their conceptual, practical, and technical difficulties. 
To overcome the difficulties, we turn to the strong connections between information and thermodynamics illuminated by Szilard's engine~\cite{Szilard:1929}, Landauer's principle~\cite{Landauer:1961}, and Bennett's information fuel tape~\cite{Bennett:1982,Feynman:1996}, and consider the concept of informational work storage~\cite{Bennett:1982,Feynman:1996,Faist:2015,Faist:2017}. 
This concept features an information-theoretic notion of work as the ability to increase the \emph{information content} of a work storage system, which is in sharp contrast with the usual concept of mechanical work storage underlying the quantum weight~\cite{Talkner:2007,Allahverdyan:2014,Talkner:2016}.
An informational work storage system based on maximally mixed qubits, capable of truly distinguishing work from heat, and well suited to thermal operations is studied and applied to the problem of single-shot work extraction~\cite{Horodecki:2013,Gemmer:2015,Wang:2017,Aberg:2013,Egloff:2015} in various extraction schemes. 
We consider the optimal situation and derive the maximum extractable work in those extraction schemes.
Our results show that, in agreement with the second law of thermodynamics, for a given system state the maximum extractable work is independent of extraction scheme. 
In doing so, we are able to resolve an inconsistency in the literature with regard to the maximum extractable work in different single-shot extraction schemes.

\section{Results}

\subsection{The framework}

We now provide a detailed presentation of our results. 
To set the stage, let us first review the formulation of thermal operations. Given a quantum system with Hamiltonian $H_S$ and a heat bath with Hamiltonian $H_B$, the allowed operations are completely positive, trace-preserving maps on a system state $\rho_S$ of the form $\mathcal{E}(\rho_S)\coloneqq\tr_B[V(\rho_S\otimes\tau_B)V^\dagger]$, or global unitary transformations $\rho_S\otimes\tau_B\xrightarrow{V}\sigma_{SB}$.
Here $\tau_B$ is the thermal state of the bath associated with $H_B$ at some temperature $T$, $\sigma_{SB}$ is the final global state of the total system, and $\mathcal{E}(\rho_S)=\tr_B(\sigma_{SB})$. 
To strictly conserve energy, the global unitaries $V$ are required to commute with the total Hamiltonian $H_S+H_B$.
It is clear that $\mathcal{E}(\tau_S)=\tau_S$, where $\tau_S$ is the thermal state of the system associated with $H_S$ at temperature $T$. Hence the thermal state $\tau_S$ is a free state, while any state other than $\tau_S$ is a resource state. 
Moreover, resource states can be classified by their free energy, which also quantifies the interconversion rate between resource states and the amount of work that can be extracted from resource states in the limit of asymptotically many copies of the quantum system~\cite{Brandao:2013}.

In classical thermodynamics the lifting of a weight a certain height is identified with the work performed by a heat engine. Analogous to this, work in the quantum regime can be thought of as the ability of a quantum system to excite a work storage system from a lower energy eigenstate to a higher one.
For convenience, we henceforth refer to any work storage system with \emph{nondegenerate} Hamiltonian $H_W$ for which work is identified with the energy difference of its energy eigenstates as a \emph{mechanical work storage system} or a weight for short.
Work extraction under thermal operations using a weight is implemented by the following energy-conserving global unitary transformations
\begin{equation}
\rho_S\otimes\tau_B\otimes|x\rangle\langle x|_W\xrightarrow{\,V\,}\sigma_{SB}\otimes|x'\rangle\langle x'|_W,\label{eq:thermalopvws}
\end{equation}
where $|x\rangle_W$ is the energy eigenstate of the weight with energy $x$ and the global unitary $V$ commutes with the total Hamiltonian $H_S+H_B+H_W$.
The work extracted from the system in a single run of the process is identified with the energy difference $w=x'-x$ of the final and the initial states of the weight. 
However, because of the quantum nature of the system and bath, the extracted work for a given system state $\rho_S$ fluctuates from run to run.
The average extracted work is defined as the average change in the weight energy over a large number of runs for many independent and identically distributed (IID) copies of the system state $\rho_S$. This is generally referred to as the IID regime in the literature.
The problem of work extraction has recently been studied in the so-called single-shot regime~\cite{Aberg:2013}, where one is  interested in the maximum amount of deterministic (fluctuation-free) work that can be extracted in each single run, apart from some failure probability $\varepsilon$~\cite{Dahlsten:2011,Aberg:2013,Horodecki:2013}.

\subsection{Difficulties with the weight}

Various idealizations of the weight with a continuous or discrete energy spectrum, and with or without a ground state, have been considered in the literature (see, e.g., Refs.~\cite{Horodecki:2013,Skrzypczyk:2014,Richens:2016,Gemmer:2015,Aberg:2014}). Despite its wide use, the weight suffers from several conceptual, practical, and technical difficulties.

(i) The energy spectrum of the weight introduces a new, yet arbitrary, energy scale $\Lambda$ to the problem that is a priori independent of the characteristic energy scale set by $kT$, where $k$ is the Boltzmann constant. 
In Ref.~\cite{Horodecki:2013}, a two-level weight (referred to by the authors as a work qubit, or wit for short) is considered. The corresponding new energy scale $\Lambda$ is the energy gap between the energy levels, and an ad hoc energy scale $\Lambda\sim kT$ is imposed by those authors. 
For the wit to function properly, the amount of work to be extracted in each single run has to be known beforehand, and the energy gap of the wit needs to be adjusted exactly to that amount. This however is problematic because the work extracted in each single run fluctuates.
In particular, as will be discussed in (iii) below, the extracted work in each single run is generally an inherently fluctuating quantity, and hence cannot be stored using just a single wit; therefore, either a collection of wits with different energy gaps or a multilevel weight with either equal or unequal energy gaps must be used.  
To circumvent this problem, the authors of Ref.~\cite{Skrzypczyk:2014} considered a weight with a continuous energy spectrum. This is tantamount to taking the apparently ad hoc limit $\Lambda\to 0$, together with the limit of an infinite number of energy levels.
However, it is noted that from a practical point of view a finite quantum system made of a few to a few tens of atoms will generally not allow for much flexibility in manipulating its energy levels to the extent of that of a weight with a continuous spectrum.

(ii) Since the initial state of the weight is \emph{not} the thermal state $\tau_W$ associated with the Hamiltonian $H_W$ at temperature $T$, the thermal state $\tau_S$ of the system is not invariant under the energy-conserving global unitary transformations in Eq.~\eqref{eq:thermalopvws}. 
In other words, by explicitly including a weight we end up with a resource theory without any free state at all. More importantly, we risk bringing in resource states for free in the guise of energy eigenstates of the weight. 
Hence independence of the extracted work on the initial state of the weight has to be imposed by hand such that the weight cannot be used as an entropy sink~\cite{Skrzypczyk:2014}, a necessary condition for defining work~\cite{Gallego:2016}.
This constraint can be fulfilled if the global unitaries $V$ commute with translations on the weight~\cite{Skrzypczyk:2014} or $[V,\Gamma_a]=0$, where $\Gamma_a$ is the translation operator on the weigh by an energy difference $a$, i.e., $\Gamma_a|x\rangle_W=|x+a\rangle_W$.
This however leads to another difficulty.

(iii) A commonly used construction for the global unitaries $V$ that satisfy the requirements $[V,\Gamma_a]=0$ and $[V,H_S+H_B+H_W]=0$ is given by~\cite{Skrzypczyk:2014,Richens:2016,Alhambra:2016,Masanes:2017}
\begin{equation}
V=\sum_i |\widetilde{i}\rangle\langle i|_{SB}\otimes\Gamma_{a_i},\label{eq:globaluV}
\end{equation}
where the state $|i\rangle_{SB}$ ($|\widetilde{i}\rangle_{SB}$) is the energy eigenstate of the system and bath with corresponding energy $E_i$ ($E_{\tilde{i}}$) at the beginning (end) of the extraction process, and $a_i=E_i-E_{\tilde{i}}$. 
In the simplest case where the local Hamiltonians at the beginning and the end of the extraction process are identical, up to a permutation, $\{|i\rangle_{SB}\}$ and $\{|\widetilde{i}\rangle_{SB}\}$ form the same energy eigenbasis of the system and bath.
It is noted that the global unitaries $V$ in Eq.~\eqref{eq:globaluV} represent a very generic class of allowed operations under which a weight in a pure initial state will generally end up in a mixed final state. Specifically, the transitions of the weight are given by  
\begin{align}
|x\rangle\langle x|_W\xrightarrow{~~}\sum_i\langle i|(\rho_S\otimes\tau_B)|i\rangle_{SB}\,|x+a_i\rangle\langle x+a_i|_W.\label{eq:weighttrans}
\end{align}
Compared with the weight transitions in Eq.~\eqref{eq:thermalopvws}, this implies that the extracted work in each single run is \emph{inherently} fluctuating.
It has been pointed out~\cite{Gallego:2016} that because there is an associated increase in the entropy of the weight, to quantify work for transitions of the form in Eq.~\eqref{eq:weighttrans} one has to properly account for the fact that the weight might act as an entropy sink. 
This observation implies that the independence of the extracted work on the initial state of the weight is a \emph{necessary} condition that the weight does not act as an entropy sink, but it is far from sufficient. 

(iv) The last difficulty concerns \emph{fluctuating} or \emph{probabilistic} work~\cite{Jarzynski:1997,Crooks:1999}, and is closely related to the previous one. Work extraction in the single-shot regime has been generalized to include transitions of the weight from a single energy level to multiple levels in a certain energy range~\cite{Gemmer:2015}. 
It is one of several recent attempts~\cite{Halpern:2015,Salek:2017,Dahlsten:2017} to link the resource-theoretic approach~\cite{Janzing:2000,Brandao:2013} with the fluctuation relation approach~\cite{Jarzynski:1997,Crooks:1999} in the single-shot regime.
Specifically, in the event of success the transitions of a weight with discrete energy levels are given by~\cite{Gemmer:2015}
\begin{equation}
|0\rangle\langle 0|_W\xrightarrow{~~}~\sum_{\mathclap{w=w_\text{min}}}^{w_\text{max}} P_W(w)|w\rangle\langle w|_W,\label{eq:genworktrans}
\end{equation}
where $w_\text{min}>0$ and the work probability distribution $P_W(w)\ne 0$ only if $w_\text{min}\le w\le w_\text{max}$. 
The careful reader may have noticed that the transitions in Eq.~\eqref{eq:genworktrans} are special cases of those in Eq.~\eqref{eq:weighttrans}.
Compared with the case of transitions to single final energy level described by Eq.~\eqref{eq:thermalopvws}, the maximum extractable work for the general case described by Eq.~\eqref{eq:genworktrans}, in the event of success, contains an \emph{additional} contribution given by~\cite{Gemmer:2015,Wang:2017} 
\begin{equation}
W^\mathrm{(add)}_\mathrm{max}=kT\log~\sum_{\mathclap{w=w_\text{min}}}^{w_\text{max}}\ee^{-\beta(w-w_\text{min})}.
\end{equation}
The additional contribution $W^\mathrm{(add)}_\mathrm{max}$ is, as it should be, independent of the initial state of the weight, provided that the energy levels of the weight are equally spaced~\cite{Wang:2017}, a necessary condition for the requirement that the global unitaries $V$ commute with translations on a weight with discrete energy levels.  
More importantly, it has been shown~\cite{Wang:2017} that because $W^\mathrm{(add)}_\mathrm{max}$ is nonnegative, sensitive to the details of the energy levels of the weight, and independent of the initial state of the system and the failure probability, it is not work-like but heat-like, and should count as heat transferred from the bath instead of work extracted from the system.
This result substantiates the observation made in Ref.~\cite{Gallego:2016} that for the transitions in Eq.~\eqref{eq:weighttrans} or \eqref{eq:genworktrans} the weight may act as an entropy sink. 
Hence the weight does not truly distinguish work from heat when the extracted work is inherently fluctuating or probabilistic.

\subsection{Informational work storage}

To overcome these difficulties, we make the following key observations. Denote the energies of the system, heat bath, and work storage system by $E_S$, $E_B$, and $E_W$, respectively. 
In each energy shell at total energy $E=E_S+E_B+E_W$, the free energy $F^E\coloneqq E-kTS^E$ of the total system is invariant under energy-conserving global unitary transformations, where $S^E$ is the (von Neumann) entropy of the total system taken in the energy shell $E$. 
In a work extraction process the free energy of the system and bath in the energy shell $E$ is expected to decrease, and hence there will be a corresponding increase of the same amount $\Delta F_W=\Delta E_W-kT\Delta S_W>0$ in the free energy of the work storage system, where $E_W$ and $S_W$ are the energy and entropy of the work storage system, respectively.
When the work storage system is a weight, this is achieved by an increase in the energy of weight $\Delta E_W=\Delta F_W$, provided that its entropy remains unchanged, i.e., $\Delta S_W=0$. 
However, if there is an increase $\Delta S_W>0$ in the entropy of the weight (e.g., the weight acts as an entropy sink), we have $\Delta E_W>\Delta F_W$; hence the transferred energy contains a part that is heat-like instead of work-like.
The crucial point to note is that the increase $\Delta F_W>0$ in the free energy of the work storage system can also be achieved by a \emph{decrease} $\Delta S_W<0$ in the entropy, while keeping the energy unchanged at the beginning and the end of the extraction process, i.e., $\Delta E_W=0$. Then we have $\Delta F_W=-kT\Delta S_W>0$; the transferred energy is associated with a decrease in the entropy and hence cannot be heat-like.

Motivated by the above observations, we resort to insights from the relations between information and work, especially the concepts of Szilard's engine~\cite{Szilard:1929}, Landauer's principle~\cite{Landauer:1961}, and Bennett's information fuel tape~\cite{Bennett:1982,Feynman:1996} (or information battery~\cite{Faist:2015,Faist:2017}). 
The information battery is a system consisting of a large number of pure qubits with a degenerate Hamiltonian, and work can be extracted from it by making the qubits maximally mixed. 
It has been used to determine the work value of information for systems with internal correlations~\cite{Dahlsten:2011}, and the minimum work cost required to carry out any logical process, e.g., a computation~\cite{Faist:2015,Faist:2017}. 
While the equivalence of the weight, work bit, and information battery for weight transitions of the form in Eq.~\eqref{eq:thermalopvws} has been established in Ref.~\cite{Brandao:2015}, we stress that the difficulties with the weight and the observations discussed above strongly indicate that this is not the case for weight transitions of the form in Eq.~\eqref{eq:genworktrans}. Hence a careful analysis of informational work storage in situations where the extracted work in each single run is inherently fluctuating is clearly warranted.

\begin{figure}[t]
\begin{center}
\includegraphics[width=4.0in]{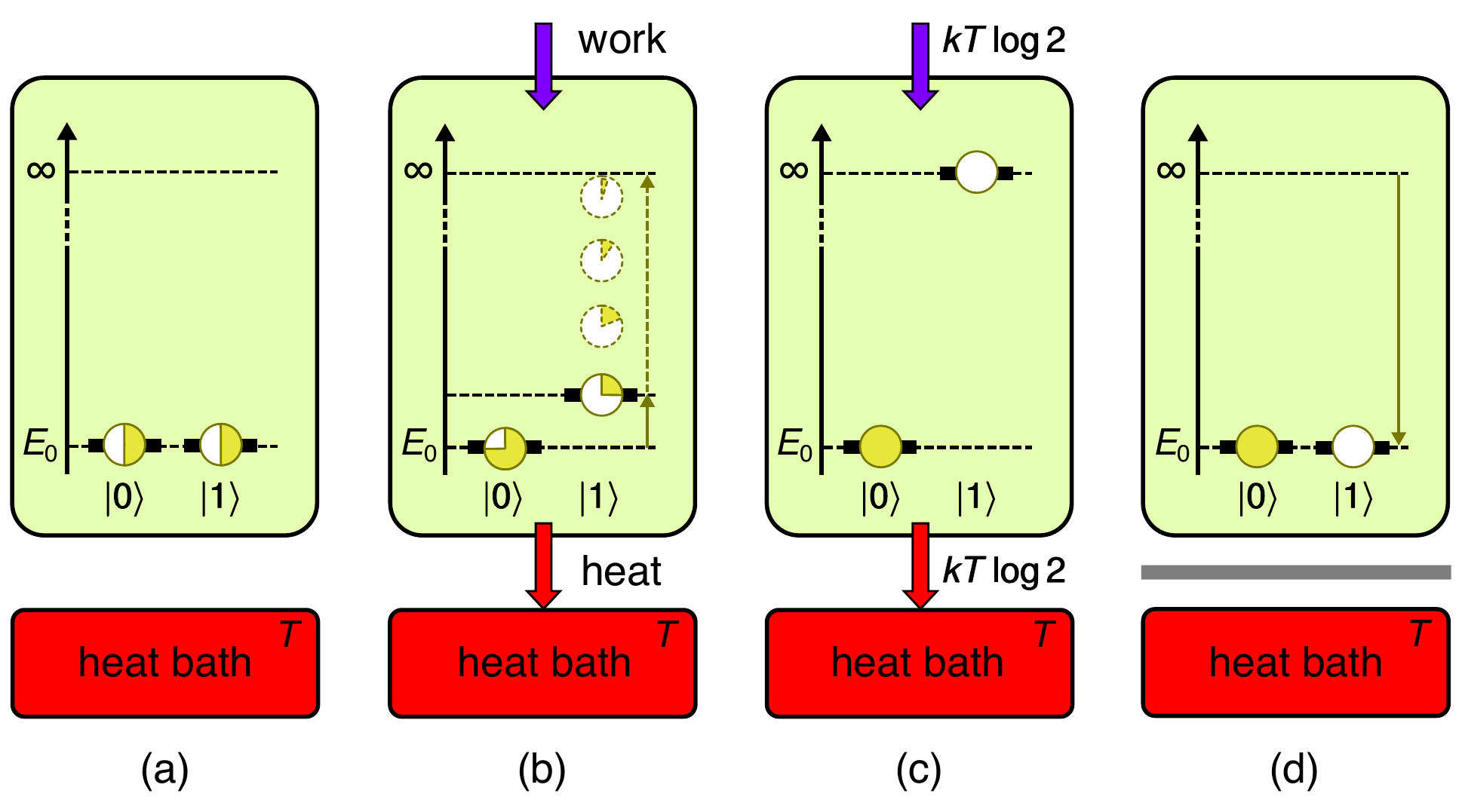}
\end{center}
\caption{Resetting a maximally mixed qubit and work storage. The solid lines represent the energy eigenstates of the qubit. 
The circles represent the probabilities of the corresponding states, and a filled circle means that the qubit is found in that state with certainty. 
(a) The qubit is initially in the state $\openone_2/2$ and coupled to the heat bath at temperature $T$. 
(b) Work extracted from the system into the driving field is used to raise the state $|1\rangle$ of the qubit in small steps while keeping the state $|0\rangle$ untouched. The qubit is allowed to thermalize with the bath after each step, resulting in heat dissipated to the bath. 
As the state $|1\rangle$ is raised, it is more likely to find the qubit in the untouched state $|0\rangle$. 
(c) In the limit of an infinite energy gap, the qubit is found in the state $|0\rangle$ with certainty, and the work done by the driving field in the quasistatic limit is $kT\log 2$.
(d) The qubit is decoupled (represented by the gray line) from the bath and the state $|1\rangle$ is lowered back to its original energy in a single step. 
There is no work cost for this process because the state $|1\rangle$ is not occupied. 
The final state of the qubit is the pure state $|0\rangle$, and $kT\log 2$ work input from the driving field is stored in the qubit.
By running the whole process in reverse, we obtain work withdrawal from a pure qubit in the state $|0\rangle$, yielding $kT\log 2$ work output to the driving field at the cost of leaving the qubit maximally mixed. Figure adapted from Ref.~\cite{delRio:2011}.}
\label{fig:figure1}
\end{figure}

According to Landauer's principle~\cite{Landauer:1961,Piechocinska:2000,Alicki:2004,delRio:2011}, the work cost of resetting a maximally mixed qubit to a pure state using an optimal (reversible) process in the presence of a heat bath at temperature $T$ is $kT\log 2$. 
The reverse process, known as the Szilard engine, implies that the possession of a qubit in a pure state and a heat bath at temperature $T$ can be used to perform $kT\log 2$ of work~\cite{Szilard:1929,Bennett:1982,Dahlsten:2011}.
The fact that together with a heat bath at temperature $T$, one bit of information (encoded in a pure degenerate qubit) can be reversibly traded for $kT\log 2$ work implies that work can be thought of as the ability to increase the \emph{information content} of a work storage system, instead of the \emph{energy content}. The work storage and withdrawal processes using a qubit as the work storage system are illustrated in Fig.~\ref{fig:figure1}.
Indeed, the energy difference of the energy eigenstates of the qubit vanishes because its Hamiltonian is degenerate at the beginning and the end of a work storage or withdrawal process.
A work storage system consisting of degenerate qubits is fundamentally different from a weight (especially a wit~\cite{Horodecki:2013}) in that the Hamiltonian of the latter is always and necessarily nondegenerate~\cite{Gallego:2016}.  
To highlight its information-theoretic character, we refer to this new kind of work storage system that consists of a large number of qubits as an \emph{informational work storage system}, or a \emph{bittery} (battery of qubits) for short.

\begin{figure}[t]
\begin{center}
\includegraphics[width=3.0in]{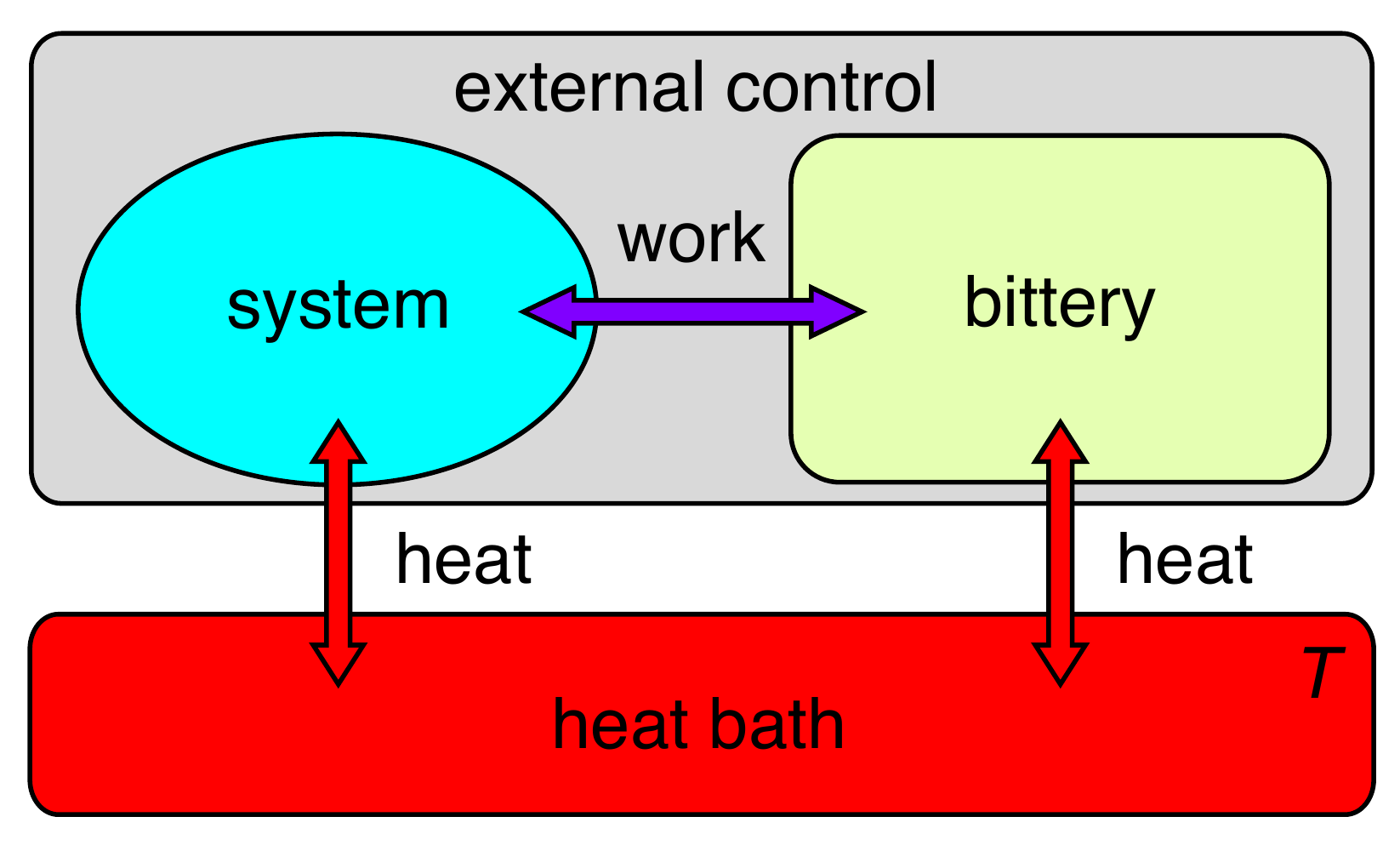}
\end{center}
\caption{Schematic setup of thermal operations using a bittery as the work storage system.}
\label{fig:figure2}
\end{figure}

Work extraction under thermal operations now involves a system, heat bath at temperature $T$, an $n$-cell bittery, and external control. A schematic setup of thermal operations using a bittery as the work storage system is shown in Fig.~\ref{fig:figure2}. 
The $n$-cell bittery consists of a certain large number $n$ of qubits, each with degenerate Hamiltonian $H_\text{qubit}$ and in the maximally mixed state $\openone_2/2$, where $\openone_d$ is the identity operator on a $d$-dimensional Hilbert space. 
To avoid possible confusion with the weight, we will denote the bittery by $A$.
Hence the $n$-cell bittery has Hamiltonian $H_A=\sum_{i=1}^n H^{i}_\text{qubit}$, initial state $2^{-n}\openone_{2^n}$, and a storage capacity of $nkT\log 2$, where $H^{i}_\text{qubit}$ is the Hamiltonian of the $i$th qubit.
The external control may perform any global unitary that strictly conserves energy.
It may also manipulate the system and bittery by (i) coupling or decoupling those systems to the heat bath such that energy (in the form of heat) is exchanged between the bath and those systems, and (ii) raising or lowering any energy level of their Hamiltonian (e.g., by tuning certain parameters) such that energy (in the form of work) is exchanged between those systems (see Fig.~\ref{fig:figure1}). 
We stress that the \emph{only} constraint imposed on the global unitaries $V$ is strict energy conservation, i.e., $[V,H_S+H_B+H_A]=0$.
At the end of a work extraction process, the state of the bittery consists of possibly $m<n$ qubits in the maximally mixed state $\openone_2/2$, while the other $n-m$ qubits are reset to the pure state $|0\rangle$, corresponding to the final state $2^{-m}\openone_{2^m}\otimes|0\rangle\langle 0|^{\otimes(n-m)}$ and an amount of $(n-m)kT\log 2$ extracted work being stored in the bittery. 
The $m$ untouched (maximally mixed) qubits can be used as an $m$-cell bittery for a subsequent work extraction process, because the $n-m$ reset (pure) qubits are decoupled from the system, bath, and external control, and not subjected to the global unitary transformation of the subsequent process.

Before applying the bittery to specific problems, we discuss the advantages of the bittery over the weight. 
In fact, all of the four difficulties with the weight that are discussed above are overcome by the advantages of the bittery.
(i) Any system consisting of a large number of maximally mixed qubits with a degenerate Hamiltonian can be used as a bittery.
The energy scale associated with the bittery is the characteristic energy scale $kT$; hence the bittery does not introduce any new energy scale to the problem. 
More importantly, experimental demonstrations of information to work conversion in the Landauer limit using various realizations of qubits in atomic and molecular systems have recently been reported~\cite{Toyabe:2010,Peterson:2016,Koskia:2014,Cottet:2017}.
(ii) Since the initial state of an $n$-cell bittery is the maximally mixed state $2^{-n}\openone_{2^n}$, or equivalently the thermal state $\tau_A$ associated with the degenerate Hamiltonian $H_A$ at temperature $T$, the $n$-cell bittery (with $n$ large and arbitrary) can always be included for free at no cost of any resource. 
The thermal state $\tau_S$ of the system is invariant under energy-conserving global unitary transformations, and hence remains a free state, even if the bittery is explicitly included. 
Consequently, there are no additional restrictions on the class of allowed energy-conserving global unitaries.
(iii) Because the maximally mixed state $2^{-n}\openone_{2^n}$ is the maximum entropy state of the $n$-cell bittery, or equivalently the thermal state $\tau_A$ is a completely passive state~\cite{Pusz:1978,Lenard:1978}, the bittery cannot be used as an entropy sink. 
(iv) The bittery has an inherent capability to truly distinguish work from heat because it costs, and only costs, work to reset maximally mixed qubits. 
The last two advantages will be demonstrated analytically in the example discussed in the next subsection.

\subsection{Application to single-shot work extraction}

Recently, the problem of extracting work from a quantum system has received much attention~\cite{Horodecki:2013,Skrzypczyk:2014,Aberg:2013,Dahlsten:2011,Egloff:2015,Gemmer:2015,Richens:2016,Wang:2017}. 
To demonstrate the advantage of the bittery, we apply it to study the work extraction problem in the single-shot regime. 
We consider the optimal situation and hence the maximum extractable work, and leave the suboptimal situations and connection with fluctuation relations~\cite{Jarzynski:1997,Crooks:1999} to future work.
Three work extraction schemes (or work contents) have been considered previously; however there is no general agreement on the maximum extractable work.
The $\varepsilon$-\emph{deterministic work} is the amount of deterministic work that can be extracted with failure probability $\varepsilon$~\cite{Aberg:2013,Horodecki:2013,Gemmer:2015,Wang:2017}, the $\varepsilon$-\emph{deterministic} $c$-\emph{bounded work} (or $(\varepsilon,c)$-deterministic work) is the $\varepsilon$-deterministic work with inherent fluctuations bounded by a given amount $c$~\cite{Aberg:2013,Gemmer:2015,Richens:2016,Wang:2017}, while the $\varepsilon$-\emph{guaranteed work} is the work that is guaranteed to be exceeded with failure probability $\varepsilon$~\cite{Egloff:2015}.
Following Refs.~\cite{Horodecki:2013,Gemmer:2015,Wang:2017}, we assume that the system and heat bath have a finite-dimensional Hilbert space and a Hamiltonian with minimum energy zero, the heat bath has large but finite energy, and the energy of the system is small compared with that of the bath.
The relevant global unitary transformations are of the form 
\begin{equation}
\eta\coloneqq\rho_S\otimes\tau_B\otimes\rho_A\xrightarrow{\,V\,}\sigma_{SB}\otimes\sigma_A\eqqcolon\eta',
\end{equation}
where $\rho_A=\cramped{\xi_A^{(n)}}$ with $\cramped{\xi_A^{(m)}}\coloneqq 2^{-m}\openone_{2^m}\otimes|0\rangle\langle0|^{\otimes(n-m)}$ for $m=0,\dots,n$, and $\sigma_A$ is the final state of the bittery. 
For simplicity~\cite{Horodecki:2013,Gemmer:2015,Wang:2017}, we consider the case where the initial system state $\rho_S$ is diagonal in the energy eigenbasis of the system and there are no correlations between the bittery and the rest in the final global state of the total system. We assume without loss of generality a trivial bittery Hamiltonian $H_A=0$, and hence zero bittery energy $E_A=0$, at the beginning and the end of the extraction process.
We use the method developed in Ref.~\cite{Wang:2017} because it not only is physically intuitive and mathematically simple, but also provides a unified approach to the three extraction schemes.
The only mathematical tool used is the Schur theorem~\cite{Marshall:2010} in majorization theory, i.e., if $\lambda_1,\ldots,\lambda_d$ and $\alpha_1,\ldots,\alpha_d$ are respectively the eigenvalues and diagonal elements of a $d\times d$ Hermitian matrix (including multiplicities), then the vector of eigenvalues $\lambda$ majorizes the vector of diagonal elements $\alpha$.

While the initial global state $\eta$ is diagonal in the global energy eigenbasis that defines the energy shells, because the total Hamiltonian $H$ has degeneracies the final global state $\eta'$ is generally not diagonal in this basis. 
However, since the global unitaries $V$ commute with the total Hamiltonian $H$, the final global state $\eta'$ is block-diagonal in the global energy eigenbasis with no off-diagonal elements between energy shells at different total energies.
Moreover, because the global unitaries $V$ are strictly energy conserving, it suffices to consider $\eta$ and $\eta'$ in the energy shell at total energy $E=E_S+E_B$, denoted respectively by $\cramped{\eta^E}$ and $\cramped{\eta'^E}$, where $E_S$ and $E_B$ are the energies of the system and bath, respectively.
We note that $\cramped{\eta^E}$ and $\cramped{\eta'^E}$ have the same eigenvalues (including multiplicities).
Denote the eigenvalues of $\cramped{\eta^E}$ and the diagonal elements of $\cramped{\eta'^E}$ in the product basis of the local energy eigenstates (which defines the energy shell $E$ and diagonalizes $\eta^E$) by $\cramped{r^E}(E_S,g_S,g_B,g_A)$ and $\cramped{s^E}(E_S,g_S,g_B,g_A)$, respectively. 
Here $g_S(E_S)=1,\ldots,M_S(E_S)$, $g_B(E_B)=1,\ldots,M_B(E_B)$, and $g_A(\cramped{\xi_A^{(m)}})=1,\ldots,2^m$ are the degeneracy indices of the system, bath, and bittery, respectively, with $M_S(E_S)$ and $M_B(E_B)$ being the corresponding multiplicities. 
According to the Schur theorem, it follows that the vector of eigenvalues $\cramped{r^E}$ majorizes the vector of diagonal elements $\cramped{s^E}$. This in turn implies~\cite{Wang:2017,Marshall:2010} 
\begin{equation}
\rk(\cramped{r^E_\varepsilon})\le\rk(\cramped{s^E_\varepsilon}),\label{eq:rkrelesew}
\end{equation}
where $\cramped{r^E_\varepsilon}$ and $\cramped{s^E_\varepsilon}$ are respectively the vectors of the \emph{largest} nonzero elements in $\cramped{r^E}$ and $\cramped{s^E}$ that add up to $(1-\varepsilon)\tr(\cramped{\eta^E})$, and rank denotes the number of those nonzero elements.
Following previous work~\cite{Horodecki:2013,Gemmer:2015,Wang:2017}, we assume that the bath multiplicity $M_B(E_B)$ at energy $E_B$ grows exponentially with the inverse temperature $\beta=1/kT$. Hence we will use the approximation $M_B(E-E_S)=M_B(E)\ee^{-\beta E_S}$ for $E\gg E_S$. 
Then we obtain~\cite{Wang:2017}
\begin{equation}
\rk(\cramped{r^E_\varepsilon})=2^n M_B(E)\sum_{\mathclap{E_S,g_S}}h_\varepsilon(E_S,g_S)\ee^{-\beta E_S},\label{eq:rankreeps}
\end{equation}
where $h_\varepsilon(E_S,g_S)$ is an indicator function that takes values in the interval $[0,1]$ and determines the fraction of the bath multiplicity to be included in the summation~\cite{Horodecki:2013,Gemmer:2015,Wang:2017}.

We first consider $\varepsilon$-deterministic extraction of $(n-m)kT \log2$ work, where $0\le m<n$. The final bittery state $\sigma_A$ is a mixture of the form
\begin{equation}
\sigma_A=\varepsilon\rho_A+(1-\varepsilon)\cramped{\xi_A^{(m)}}.
\end{equation}
Henceforth, by failure of work extraction we mean in the strict sense of the term that there is no work being extracted from the system at all~\cite{Wang:2017}. 
Otherwise, an arbitrary amount of work could still be extracted in a failed work extraction, thus rendering the energy transfer unpredictable and more heat-like than work-like.
In the event of success the entropy change of the bittery is 
\begin{equation}
\Delta S_A=S(\cramped{\xi_A^{(m)}})-S(\rho_A)=(m-n)\log 2<0,
\end{equation}
where $S(\rho)=-\tr(\rho\log\rho)$ is the von Neumann entropy, and hence the entropy of the bittery never increases. 
The maximum extractable work can be found using the fact that $\rk(\cramped{s^E_\varepsilon})$ is upper bounded~\cite{Wang:2017}. 
Since the elements in $\cramped{s^E}$ that correspond to $\cramped{\xi_A^{(m)}}$ scale as $2^{-m}$, to ensure that the extractable work is as large as possible, we require the upper bound being as stringent as possible.
The most stringent upper bound on $\rk(\cramped{s^E_\varepsilon})$ is
\begin{align}
\rk(\cramped{s^E_\varepsilon})&\le
\dim(\sigma_{SB}^E)\rk(\cramped{\xi_A^{(m)}})\nn\\
&=2^m\sum_{\mathclap{E_S,g_S}}M_B(E-E_S)\nn\\
&=2^m Z_S M_B(E),\label{eq:ranksee}
\end{align}
where $\sigma_{SB}^E$ denotes $\sigma_{SB}$ in the energy shell $E$, $\rk(\cramped{\xi_A^{(m)}})$ denotes the number of nonzero eigenvalues of $\xi_A^{(m)}$, and $Z_S=\tr(\ee^{-\beta H_S})$ is the partition function of the system at temperature $T$.
From Eqs.~\eqref{eq:rkrelesew}, \eqref{eq:rankreeps}, and \eqref{eq:ranksee}, we find the maximum $\varepsilon$-deterministic work to be given by
\begin{equation}
W^\mathrm{max}_\varepsilon=F^\mathrm{min}_\varepsilon(\rho_S)-F(\tau_S),
\end{equation}
where $F^\mathrm{min}_\varepsilon(\rho_S)$ is the $\varepsilon$-smooth min-free energy of the system state $\rho_S$ given by~\cite{Aberg:2013,Horodecki:2013,Gemmer:2015,Wang:2017} 
\begin{equation}
F^\mathrm{min}_\varepsilon(\rho_S)=-kT\log\sum_{\mathclap{E_S,g_S}}h_\varepsilon(E_S,g_S)\ee^{-\beta E_S},
\end{equation}
and $F(\tau_S)=-kT\log Z_S$ is the standard free energy of the system.
Note that $W^\mathrm{max}_\varepsilon$ is independent of the energy shell $E$ because the trivial dependence of $M_B(E)$ on $E$ cancels out.
Hence $W^\mathrm{max}_\varepsilon$ is indeed the maximum $\varepsilon$-deterministic work.

Next, we consider $(\varepsilon,c)$-deterministic extraction of $(n-l)kT\log2$ work, where $m_1\le l\le m_2$ with $0\le m_1,m_2<n$ integers and $c=m_2-m_1$ fixed. The corresponding final bittery state $\sigma_A$ is of the form
\begin{equation}
\sigma_A=\varepsilon\rho_A+(1-\varepsilon)\sum_{\mathclap{l=m_1}}^{m_2} P_A(l)\cramped{\xi_A^{(l)}},\label{eq:sigmaaepscb}
\end{equation}
where the work distribution $P_A(l)\ne 0$ only if $m_1\le l\le m_2$.
As can be seen from Eq.~\eqref{eq:genworktrans}, this is exactly the case of general work extraction considered in Ref.~\cite{Gemmer:2015}, where, as discussed above, the extracted work in each single run is inherently fluctuating.
In the event of success the entropy change of the bittery is
\begin{align}
\Delta S_A&=S\left(\sum_{l=m_1}^{m_2} P_A(l)\cramped{\xi_A^{(l)}}\right)-S(\rho_A)\nn\\
&<\sum_{l=m_1}^{m_2} P_A(l)[l\log 2-\log P_A(l)]-n\log 2\nn\\
&\le\log(m_2-m_1+1)+(m_2-n)\log 2,
\end{align} 
where the first inequality is a strict one because the bittery states $\cramped{\xi_A^{(l)}}$ generally have support on nonorthogonal subspaces~\cite{Nielsen:2010}.
It follows that $\Delta S_A<0$ provided $n-m_2\ge\log_2(m_2-m_1+1)$. This condition is satisfied because to make sense of work with bounded fluctuations, the extracted work has to be at least as large as the fluctuation bound, i.e., $n-m_2\ge m_2-m_1$, thus implying $n-m_2\ge m_2-m_1\ge\log_2(m_2-m_1+1)$, where the second inequality is valid for all integers $m_2-m_1\ge 0$.
The most stringent upper bound on $\rk(\cramped{s^E_\varepsilon})$ is
\begin{align}
\rk(\cramped{s^E_\varepsilon})&\le
\dim(\sigma_{SB}^E)\smashoperator[r]{\min_{m_1\le l\le m_2}}~\rk(\cramped{\xi_A^{(l)}})\nn\\
&=2^{m_1} Z_S M_B(E),
\end{align}
from which the maximum $(\varepsilon,c)$-deterministic work is found to be given by $W^\mathrm{max}_\varepsilon$ and independent of $c$.
Thus, the inherent fluctuations in the extracted work do \emph{not} affect the maximum extractable work.

Finally, for $\varepsilon$-guaranteed extraction of ${(n-m)}kT\log2$ work we have
\begin{equation}
\sigma_A=\varepsilon\rho_A+(1-\varepsilon)\sum_{l=0}^m P_A(l)\cramped{\xi_A^{(l)}},\label{eq:sigmaagw}
\end{equation}
where $P_A(l)\ne 0$ only if $0\le l\le m$. 
Clearly, Eq.~\eqref{eq:sigmaagw} is a special case of Eq.~\eqref{eq:sigmaaepscb} with $m_1=0$ and $m_2=m$. We thus obtain the most stringent bound
\begin{equation}
\rk(\cramped{s^E_\varepsilon})\le\dim(\sigma_{SB}^E)\rk(\cramped{\xi_A^{(0)}})=Z_S M_B(E),
\end{equation}
and hence the maximum $\varepsilon$-guaranteed work is still given by $W^\mathrm{max}_\varepsilon$.
Therefore, in accordance with the second law of thermodynamics, for a given system state the maximum extractable work is independent of extraction scheme. 
More interestingly, this is akin to a family of second laws~\cite{Brandao:2015}, one for each value of $\varepsilon$, in the single-shot regime in that all optimal work extraction schemes with failure probability $\varepsilon$ will extract the same maximum amount of work.
Our result agrees with those of previous work for $\varepsilon$-deterministic work~\cite{Horodecki:2013,Aberg:2013,Gemmer:2015,Wang:2017} and $\varepsilon$-guaranteed work~\cite{Egloff:2015}. 
It is noted that the results of Refs.~\cite{Aberg:2013,Egloff:2015} are obtained in a distinct setup without explicitly including a weight.
However, our result differs from those for $(\varepsilon,c)$-deterministic work~\cite{Gemmer:2015,Richens:2016}.
We hasten to stress that because a weight is explicitly included in Refs.~\cite{Gemmer:2015,Richens:2016} and because the $c$-bounded work considered there corresponds to the work for the transitions in Eq.~\eqref{eq:genworktrans}, it is likely that the weight acts as an entropy sink and hence the energy transferred to the weight is not truly work-like.

\section{Conclusions}

We studied the suitability of using quantum weights as work storage systems in thermal operations and their capability of truly distinguishing work from heat. 
A detailed analysis of the conceptual, practical, and technical problems with the quantum weights was presented. 
Motivated by the relations between information and work dictated by Szilard's engine, Landauer's principle, and Bennett's information battery, we considered the concept of informational work storage as a solution to those problems.
This concept features an information-theoretic notion of work as the ability to increase the information content of a work storage system.
An informational work storage system based on maximally mixed qubits, not acting as an entropy sink, and capable of truly distinguishing work from heat was studied and applied to the problem of extracting work from a quantum system in the single-shot regime. 
While three work extraction schemes have been considered in the literature, there is however no general agreement on the maximum extractable work. 
The disagreement signals both a conceptual inconsistency and a violation of the second law of thermodynamics. 
Using maximally mixed qubits as the informational work storage system in thermal operations, we showed that the maximum extractable work for any given system state is the same for the three extraction schemes considered previously. 
The result is in agreement with the second law of thermodynamics and helps resolve the inconsistency. We consider this an important contribution of this work.
Since the informational work storage system has the  ability to deal with fluctuating work, a natural extension of this work is to study suboptimal work extraction and the connection with fluctuation relations in the single-shot regime~\cite{Halpern:2015,Salek:2017,Dahlsten:2017}.  

Recently, thermodynamic resource theory has been generalized to situations where the system can exchange not only energy with a heat bath, but also particles with a particle reservoir~\cite{Halpern:2016,Halpern:2018}. The formalism of the generalized thermodynamic resource theory is analogous to that of thermal operations. In particular, the energy-conserving global unitaries are required to conserve particle number. 
The concept of informational work storage, the use of maximally mixed qubits as the informational work storage system, and the method for deriving maximum extractable work that we presented in this article are very general in nature, and hence not specific to thermal operations. 
Moreover, since the informational work storage system exchanges only energy with the system and heat bath, it is conceivable that it would be well suited to the generalized thermodynamic resource theory. 
Further study along this line will be the subject of future work.

\vspace{3ex}

\acknowledgments{I would like to thank C.N.\ Leung for a reading of an early version of the manuscript and comments.}

\end{document}